\documentclass[aps,pra,twocolumn,showpacs,floatfix,superscriptaddress]{revtex4-1}
\usepackage{amsmath,amssymb,amsfonts,bbm,graphicx,hyperref,times}
\usepackage{lmodern}

\usepackage[T1]{fontenc}
\usepackage{color}
\usepackage{float}
\usepackage{amssymb}
\usepackage{graphicx}
\usepackage{esint}
\usepackage{hyperref}

\providecommand{\tabularnewline}{\\}

\def\Tr{\hbox{Tr}} \def\sigmaCM{\boldsymbol{\sigma}}

\begin{document}

\title{Continuous-variable phase-estimation with unitary and random linear
disturbance}
\author{Douglas Delgado de Souza}
\affiliation{Instituto de Fisica Gleb Wataghin, Universidade Estadual de Campinas, 13083-970, Campinas, SP, Brazil}
\affiliation{QOLS, Blackett Laboratory, Imperial College London, London SW7 2BW, UK}
\author{Marco G. Genoni}
\affiliation{Department of Physics \& Astronomy, University College London, 
Gower Street, London WC1E 6BT, United Kingdom}
\author{M. S. Kim}
\affiliation{QOLS, Blackett Laboratory, Imperial College London, London SW7 2BW, UK}

\begin{abstract}
We address the problem of continuous-variable quantum phase estimation  in the presence of linear disturbance at the Hamiltonian level, by means of Gaussian probe states. In particular we discuss both unitary and random disturbance, by considering the parameter which characterizes the {\em unwanted} linear term present in the Hamiltonian as fixed ({\em unitary disturbance}) 
or random with a given probability distribution ({\em random disturbance}). 
We derive the optimal input Gaussian states at fixed energy, maximizing the quantum Fisher information over the squeezing angle and the squeezing energy fraction, and we discuss the scaling of the quantum Fisher information in terms of the output number of photons $n_{out}$. We observe that in the case 
of unitary disturbance the optimal state is a squeezed vacuum state and the quadratic scaling is 
conserved. As regards the random disturbance, we observe that the optimal squeezing fraction 
may not be equal to one, and, for any non-zero value of the noise parameter, the quantum Fisher information scales linearly with the average number of photons. We finally discuss the performance
of homodyne measurement, comparing the achievable precision with the ultimate limit posed
by the quantum Cram\'er-Rao bound.
\end{abstract}
\maketitle

\section{Introduction}
The usefulness of the non-classical features of quantum mechanics to perform ultra-precise 
measurement beyond the classical limit has recently indicated quantum metrology 
as one of the most promising quantum technologies \cite{GiovaNatPhot}.
Phase-estimation is the paradigmatic example of estimation problem, and the possible quantum enhancement has been widely studied both theoretically and experimentally \cite{HBstates,Bollinger96,Kok02,GiovaScience,GiovaPRL,Nagata07,Yonezawa12}.
More recently, the detrimental effect of the interaction with the 
environment has been deeply studied, showing that the ultimate quantum
limit is likely to be lost in the presence of noise \cite{Rafal12,Escher11}.
Typically, one considers the situation where
the phase-rotation is performed on the initially pure probe state, and
the noisy channel is applied afterwards on the  encoded states.
By following this approach, the role of loss \cite{Rafal09,Kacpr09,Crowley14,Knysh11} 
and phase-diffusion \cite{Brivio10,Genoni11,Genoni12,Escher12,Knysh13,Knysh14,Mihai14} 
have been investigated in great detail.
However one can also consider the case where an unwanted, 
but known and fixed term is present in the Hamiltonian generating 
the  phase rotation, influencing the estimation process. This problem
has been studied for the first time in \cite{de_pasquale_quantum_2013} 
by De Pasquale and coauthors who addressed the problem as 
{\em unitary disturbance}.
One can then consider a more general, and probably more realistic case,
where the disturbance parameter characterizing the additional term in the Hamiltonian 
is a random
variable, distributed according to a known probability distribution. We will refer to this case 
as {\em random disturbance}. \\
In this paper we consider continuous-variable phase estimation with both 
unitary and random disturbance, where the additional term in the 
Hamiltonian is linear in the bosonic operators describing the quantum
system under exam. Ideal phase-estimation with Gaussian probe states
has been firstly investigated in \cite{monras_optimal_2006}; it was demonstrated
that squeezed states are optimal, and that the 
corresponding quantum Fisher information (QFI) scales quadratically with
the average number of photons, showing the enhancement compared 
to the classical linear scaling obtainable with coherent states.
Here we will derive the QFI both for unitary and random linear disturbance,
finding the influence of the noise parameters on the optimal input Gaussian states,
and on the corresponding scaling between the QFI and the output states'
average number of photons. The manuscript is organized as follows: in Sec. \ref{s:QET}
we introduce quantum estimation theory, along with the formulas for the 
QFI in the case of unitary disturbance and for generic Gaussian states.
In Sec. \ref{s:ULD} and \ref{s:RLD} we present the result concerning the optimal
QFI and the optimal probe states for respectively the unitary  and 
random disturbance case. At the end of both sections we discuss the precision
achievable via homodyne detection in the relevant cases, 
comparing it with the ultimate bounds just derived. Sec. \ref{s:concl} ends the
paper with some concluding remarks.

\section{Quantum Estimation Theory}\label{s:QET}
Let us consider a family of quantum states $\varrho_\phi$, where $\phi$ is the 
parameter one wants to estimate. A measurement, parametrized
by positive operator valued measure (POVM) operators $\{\Pi_x\}$, 
can be fully described by means of the conditional probability
distribution $p(x|\phi) = \hbox{Tr}[ \varrho_\phi \Pi_x]$.
The corresponding precision on the estimation 
of the parameter $\phi$, by means of the measurement
$\{\Pi_x\}$ is bounded as
\begin{align}
\delta\phi \geq \frac{1}{M F(\phi)}\:, \label{eq:CCRB}
\end{align}
where $M$ is the number of measurements and
\begin{align}
F(\phi) = \int dx \: p(x|\phi) \left( \partial_\phi \log p(x|\phi) \right)^2 \label{eq:FI}
\end{align}
is the classical Fisher information (FI). The inequality (\ref{eq:CCRB})
is called {\em Cram\'er Rao bound} (CRB), and it holds for
every classical estimation problem described by a conditional
probability $p(x|\phi)$. The bound is  achievable
by means of maximum likelihood and bayesian estimators in the
limit of large number of measurements.\\
By considering the quantum case, and defining the symmetric logarithmic 
derivative operator $L_\phi$ by means of the equation
\begin{align}
2 \: \partial_\phi \varrho_\phi = L_\phi \varrho_\phi + \varrho_\phi L_\phi \:,
\end{align}
it is possible to demonstrate that the FI, for any POVM, 
is bounded from above as 
\begin{align}
F(\phi) \leq H(\phi) \label{eq:QCRB1}
\end{align}
where $H(\phi) = \hbox{Tr}[\varrho_\phi L_\phi^2]$ is the QFI \cite{HelstromBook,MatteoIJQI}.
This inequality leads to the quantum Cram\'er-Rao bound (QCRB)
\begin{align}
\delta\phi \geq \frac{1}{M H(\phi)}\:. \label{eq:QCRB2}
\end{align}
It can be demonstrated that this bound is always in principle
achievable, that is, there is always a POVM whose 
corresponding classical FI is equal to the QFI. 
It is then clear by observing Eq. (\ref{eq:QCRB2})
that a larger value of the QFI corresponds to a higher precision
achievable by means of the encoded state $\varrho_\phi$.
\subsection{Quantum estimation with unitary disturbance}
Let us first consider the case of quantum estimation of unitary parameters. 
If the parameter is encoded via a unitary operation with a hermitian
generator $G$,  
\begin{align}
\varrho_\phi = U_\phi \varrho_0 U_\phi^\dag \:\:, \: U_\phi = e^{-i \phi G}
\end{align}
and the probe state $\varrho_0=|\psi_0\rangle\langle \psi_0|$ is pure, 
the QFI can be easily evaluated as 
\begin{align}
H(\phi) = 4 \Delta^2 G = 4 \left( \langle \psi_0|G^2|\psi_0\rangle -\langle \psi_0|G|\psi_0\rangle^2\right)\:.
\end{align}
One can rather consider the case where an additional term is present
in the generator of the unitary interaction encoding the parameter $\phi$,
{\em i.e.} when the unitary operation reads
\begin{align}
U_{\phi,\eta} = \exp\{ -i H(\phi,\eta)\} := \exp \{ - i (\phi G + \eta A)\}
\label{eq:uld}
\end{align}
where $\eta \in \mathbbm{R}$ is a fixed noise parameter and $A$ is the
additional disturbance (hermitian) operator.
In \cite{de_pasquale_quantum_2013}, the authors
addressed this general problem and derived the 
following formula to calculate the QFI:
\begin{align}
H(\phi) = 4 \:\Delta^2 \bar{G}(\phi,\eta) \label{eq:QFI_UD}
\end{align}
where 
\begin{align}
\bar{G}(\phi,\eta) = \int_0^1 dt \: e^{i H(\phi,\eta)t} \:G\: e^{-i H(\phi,\eta) t} \:.
\end{align}
\subsection{Quantum estimation with Gaussian states} \label{s:estiGauss}
Let us consider a quantum system described by bosonic
operators $[a,a^\dag]=\mathbbm{1}$. A quantum state
$\varrho$ can be fully described by its characteristic function
$\chi[\varrho](\alpha) = \hbox{Tr}[\varrho D(\alpha)]$, where
$D(\alpha) = \exp\{\alpha a^\dag - \alpha^* a\}$ is the 
displacement operator in phase-space. If the characteristic
function $\chi[\varrho](\alpha)$ is a Gaussian function, 
the state is said to be Gaussian \cite{GaussAOP}. By defining the quadrature
operators vector ${\bf X}=(Q,P)^T$, where
\begin{align}
Q=a+a^\dag  \:\:\: P =-i(a-a^\dag) \:, \label{eq:QP}
\end{align}
the Gaussian quantum state can be fully described by the corresponding
average values $\bar{{\bf X}}$ and the covariance matrix
$\boldsymbol{\sigma}$, defined as 
\begin{align}
\bar{X}_j  &= \langle \psi_0 | X_j |\psi_0 \rangle \:, \\
\sigma_{jk} &= \frac12 \langle \psi_0 |X_j X_k + X_k X_j  |\psi_0 \rangle - 
\bar{X}_j \bar{X}_k \:.
\end{align}
If we are considering an estimation problem where the quantum
state $\varrho_\phi$ is Gaussian, one can derive closed formulas
for the QFI in terms of the vector $\bar{\bf X}_\phi$ and the 
matrix $\boldsymbol{\sigma}_\phi$ only \cite{MonrasGauss,PinelGauss},
obtaining
\begin{align}
H(\phi) = \frac12 \frac{\hbox{Tr}[(\sigmaCM_\phi^{-1}\sigmaCM_\phi^\prime)^2]}{1+\mu_\phi^2}
+ 2 \frac{(\mu_\phi^{\prime})^2}{1-\mu_\phi^4} + {\bf \Delta} \bar{\bf X}_\phi^{\prime {\sf T}}\sigmaCM_\phi^{-1} {\bf \Delta} \bar{\bf X}_\phi^{\prime } \:. \label{eq:GaussQFI}
\end{align}
In the formula  $\mu_\phi=\Tr[\varrho_\phi^2] = 1/\sqrt{\det[\sigmaCM]}$ represents the purity of the state,
primed quantities corresponds to derivative with respect to the parameter $\phi$, except
for ${\bf \Delta} \bar{\bf X}_\phi^{\prime}$ which is defined as
\begin{align}
{\bf \Delta} \bar{\bf X}_\phi^{\prime} = \left.\frac{d (\bar{\bf X}_{\phi+\epsilon} - \bar{\bf X}_{\phi})}{d\epsilon} \right|_{\epsilon=0} \:.
\end{align}
This expression will be extremely useful to analytically calculate the QFI for the 
{\em random linear disturbance} in Sec. \ref{s:RLD}.
\section{Unitary Linear Disturbance} \label{s:ULD}
\begin{figure}[h!]
\begin{centering}
\includegraphics[width=0.95\columnwidth]{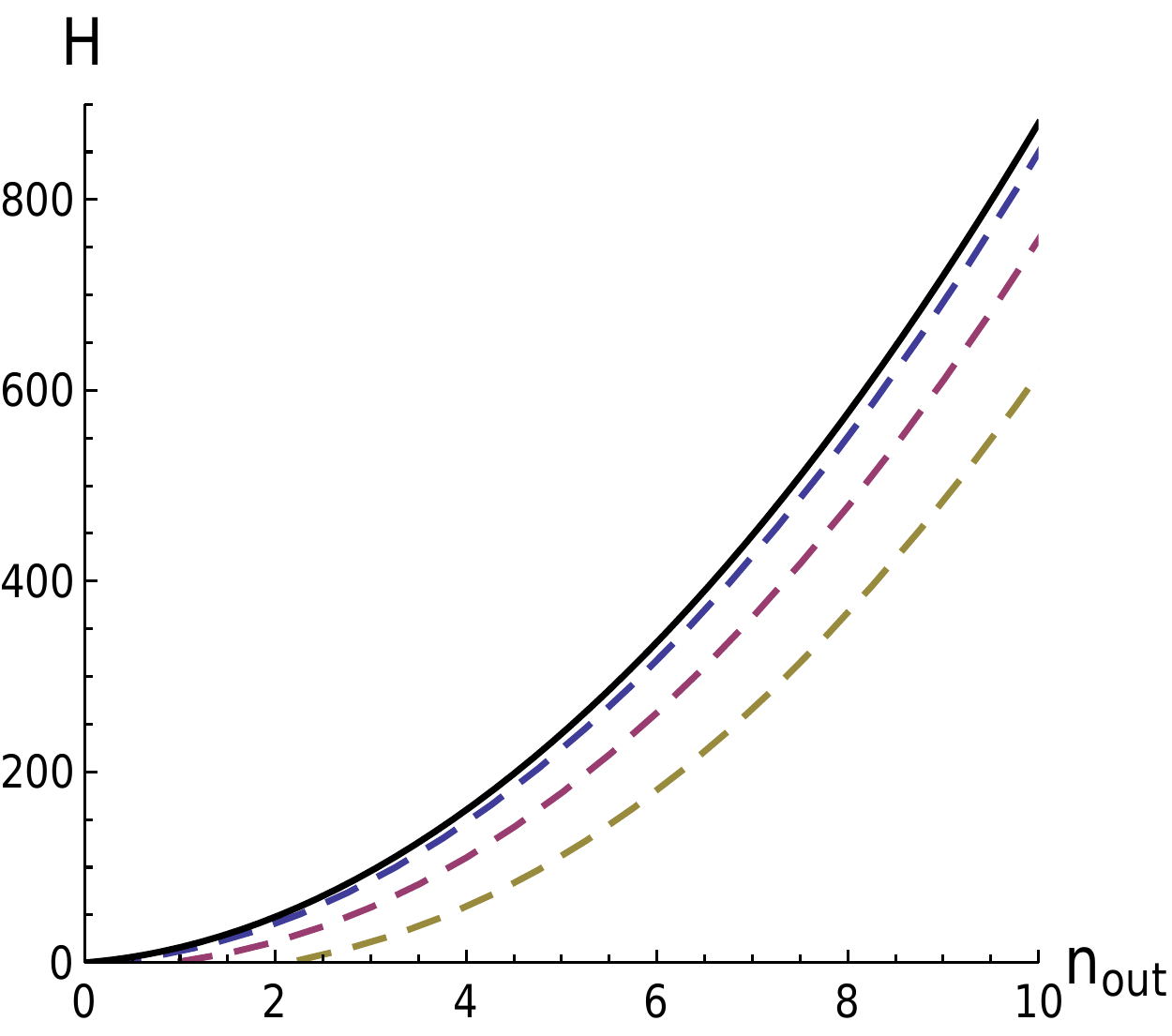}
\end{centering}
\caption{\label{f:ULD} QFI, $H$, as a function of the output number of photons $n_{\sf out}$. Solid line: noiseless estimation ($\eta=0$); dashed-line: estimation with unitary disturbance for different values
of $\eta$. From top to bottom: $\eta=\{0.5,1,1.5\}$.}
\end{figure}

In the following we study the problem of phase-estimation
with unitary linear disturbance, by considering the
unitary operator in Eq. (\ref{eq:uld}) where 
the generator and the disturbance operator
read respectively 
\begin{align}
G = a^\dag a \:, \qquad
A = Q=a+a^\dag \:,
\end{align}
such that, 
\begin{align}
U_{\phi,\eta} = \exp\{-i (\phi a^\dag a + \eta (a+a^\dag) )\}.
\end{align}
Its effect is a phase rotation accompanied by a 
displacement in phase-space; however, as the 
operators $G$ and $D$ do not commute, the two
effects cannot be separated. We are interested in
small fluctuations around a given value of the phase,
and in particular we will discuss the results regarding the 
estimation precision for $\phi=0$.\\
We consider a pure probe state 
$\varrho=|\psi_0\rangle\langle \psi_0|$, where
$$|\psi_0\rangle = D(\alpha) S(\xi) |0\rangle$$ is a generic
single-mode pure Gaussian state, $S(\xi)=
\exp\{\xi a^2 - \xi^*a^{\dag 2} \}$ is the squeezing operator, $\xi=r e^{i\theta}$, and
 $\left\{ \alpha,r,\theta\right\} \in\mathbb{R}$.
As we consider the effect of phase-rotation over a single-mode state, 
we are implying that we already have, as an implicit resource, a 
reference beam (typically a strong coherent state), 
such that the phase-rotation is well defined with respect to this reference, 
and relative phases between terms with different photon number
become observable. 
We  then focus our attention on the behavior of the QFI as a 
function of the energy of the input state $|\psi_0\rangle$.
A useful re-parametrization corresponds to considering the three
parameters $\{ n_0, \beta, \theta\} \in \mathbbm{R}$, where
\begin{align}
n_0 = \langle \psi_0|a^\dag a |\psi_0\rangle = \alpha^2 + \sinh^2 r 
\end{align}
is the average number of photons in the probe state, and
\begin{align}
\beta = \frac{\sinh^2 r}{n_0} 
\end{align}
is its squeezing fraction (for $\beta=0$ the probe state is a 
coherent state, while for $\beta=1$ it is a squeezed vacuum state). 
Our main goal is to derive the maximum value of QFI for an input 
Gaussian state at fixed number of photons $n_0$, by 
optimizing it over the parameters $\theta$ and $\beta$. 
The QFI for a generic  state $U_{\phi\eta}|\psi_0\rangle$ 
can be evaluated analytically by means of either the formula
in Eq. (\ref{eq:QFI_UD}) or the one in Eq. (\ref{eq:GaussQFI}). 
Its maximization over the squeezing angle, yields $\theta_{\sf opt}=0$,
which corresponds to squeezing of the $Q$ quadrature, while the
numerical optimization over the squeezing fraction yields $\beta_{\sf opt}=1$,
indicating that the optimal strategy is to use all the photons to prepare
a squeezed vacuum state. These results can be
 understood by observing that, for small values of  $\phi$, the
 evolution $U_{\phi,\eta}$ corresponds to a phase-space displacement 
 along the negative $P$-axis, followed by a phase-rotation depending
 on $\phi$. Amplitude squeezing thus represents the best resource in order to 
 detect the parameter $\phi$. In particular the maximized QFI for $\phi=0$ 
 reads
 \begin{align}
 H=8n_0(n_0+1)+\left(2n_0+2\sqrt{n_0(n_0+1)}+1\right)\eta^{2} \:.
 \end{align}
 As expected, in the undisturbed case of $\eta=0$, one re-obtains
 the result derived by Monras in \cite{monras_optimal_2006}. We also
 observe that the second positive term implies that the estimation 
 is improved over the case of no disturbance. This apparently
 counterintuitive result can be understood by taking into account the fact that
 the additional term in the Hamiltonian does actually increase the 
 output average number of photons, which reads 
 $n_{\sf out}=n_0 + \eta^2$. As the additional energy is used as 
 a resource for estimating the phase $\phi$, it is more interesting to consider
 the behavior of the QFI $H$ as a function of $n_{\sf out}$, in 
 order to fairly discuss the scaling as a function of the 
 number of photons. As we can see in Fig. \ref{f:ULD}, at fixed values
 of $n_{\sf out}$, the QFI takes smaller values by increasing the
 disturbance parameter $\eta$. Remarkably, we also observe that
 the {\em non-classical} quadratic scaling is still preserved, showing
 how the non-classical resource (squeezing) is fundamental to
 get the ultimate estimation precision. 
\subsection*{Performance of homodyne detection}
\begin{figure}[H]
\begin{centering}
{\bf (a)}\hskip4cm{\bf (b)}\\
\begin{tabular}{cc}
\includegraphics[width=0.50\columnwidth]{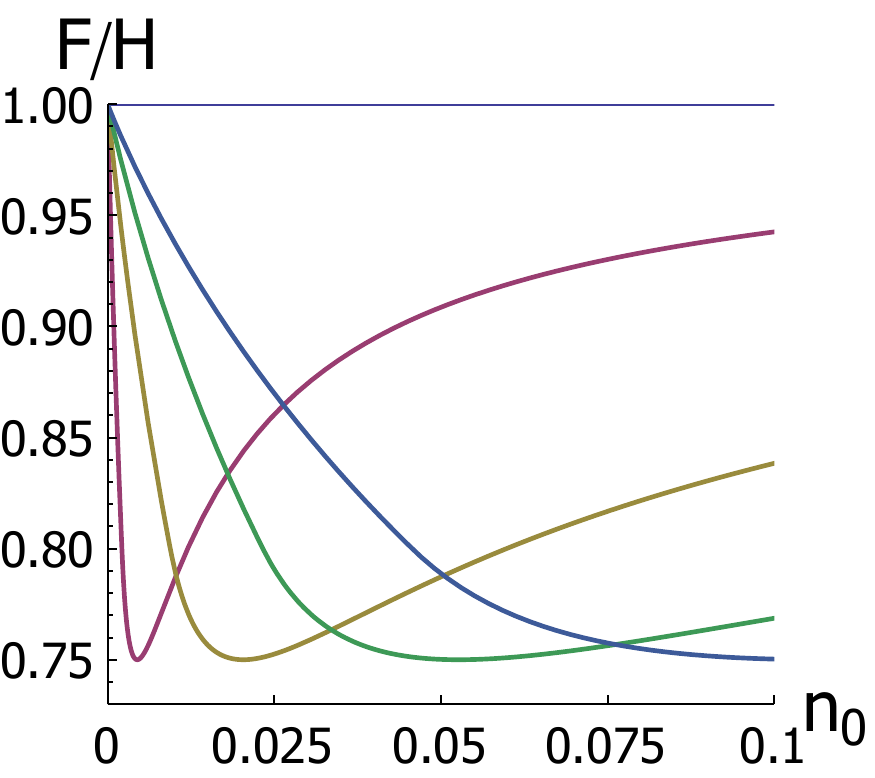} & \includegraphics[width=0.46\columnwidth]{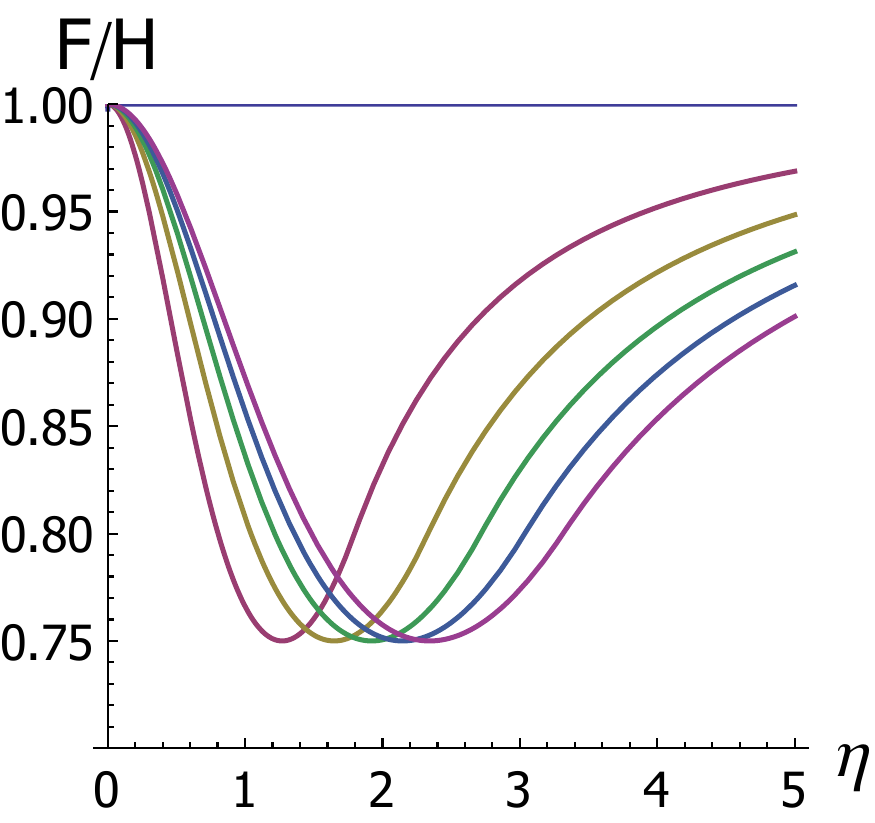}
\tabularnewline
\end{tabular}
\par\end{centering}
\caption{\label{fig:FoverQ_Unitary_disturbance} (Color online) Ratio between
the homodyne detection FI, $F$, and the corresponding
QFI, $H$. (a) $F/H$ as function of the number of photons of the probe for different
values of disturbance $\eta$; using the right ends of the curves as
reference, from top to bottom: $\eta=\{0.0,\,0.25,\,0.50,\,0.75,\,1.0\}$.\\
(b) $F/H$ as function of the disturbance parameter for different
values of the average number of photons of the probe; 
from top to bottom: $n_0=\{0, 0.2, 0.4, 0.6, 0.8, 1.0\}$.
}
\end{figure}

We now examine the suitability of homodyne detection to estimate 
the phase in the case of unitary linear disturbance. 
Homodyne detection corresponds to projection over the eigenstates
$\{\Pi_\omega(x)= |x_\omega\rangle\langle x_\omega|\}$ of the generalized quadrature
operator $X_\omega = a e^{i\omega} + a^\dag e^{-i \omega}$. The 
corresponding Fisher information is evaluated as in Eq. 
(\ref{eq:FI}), where the conditional probability reads
\begin{align}
p(x|\phi) = | \langle x_\omega | U_{\phi,\eta} |\psi_0\rangle |^2 \:.
\end{align}
We consider a squeezed vacuum state, which, as derived above, 
is the optimal probe in the case $\phi\approx 0$. 
We will then optimize the FI over the homodyne angle $\omega$, and 
compare the result to the QFI by evaluating the ratio $F/H$.
In \cite{monras_optimal_2006} it
was shown that, in the case of no disturbance, homodyne detection
is optimal as the corresponding FI is equal to the QFI.
As we can see in Fig. \ref{fig:FoverQ_Unitary_disturbance} (a),
at fixed disturbance parameter $\eta$, homodyne detection
ceases to be optimal for probes with nonzero photon number,
but it is nearly optimal when the probe is very
weak ($n_0 \approx0$) or very strong ($n_0\gg \eta^2$). The near-to-optimality
for a weak probe can be understood by observing the fact that,
the output state is basically a coherent state (due to the disturbance
in the unitary operator), and homodyne detection is optimal
for phase-estimation with a coherent state. 
Regarding input states with a large average photon number, the
disturbance can be considered as a small perturbation, and the noiseless 
optimality is recovered.\\
A symmetric description holds for the behavior of the ratio $F/H$ 
for fixed values of the average input photon number $n_0$ and as a function of the 
disturbance parameter $\eta$, which is plotted in Fig. 
\ref{fig:FoverQ_Unitary_disturbance} (right). Homodyne detection
is indeed optimal for small and large values of $\eta$ (compared to
the input photon number), corresponding in this case respectively to the
situation where the disturbance can be considered as a small
perturbation, or when the output state resembles a coherent
state.\\
In Figs. \ref{fig:FoverQ_Unitary_disturbance} (a) and (b) we also note that the ratio has a
minimum which is always equal to $(F/H)_{\sf min} = 3/4$, showing the
overall efficiency of homodyne detection 
for the whole range of the parameters.\\
\section{Random Linear Disturbance} \label{s:RLD}
\begin{figure}[h]
\begin{centering}
\begin{tabular}{cc}
\includegraphics[height=6.5cm]{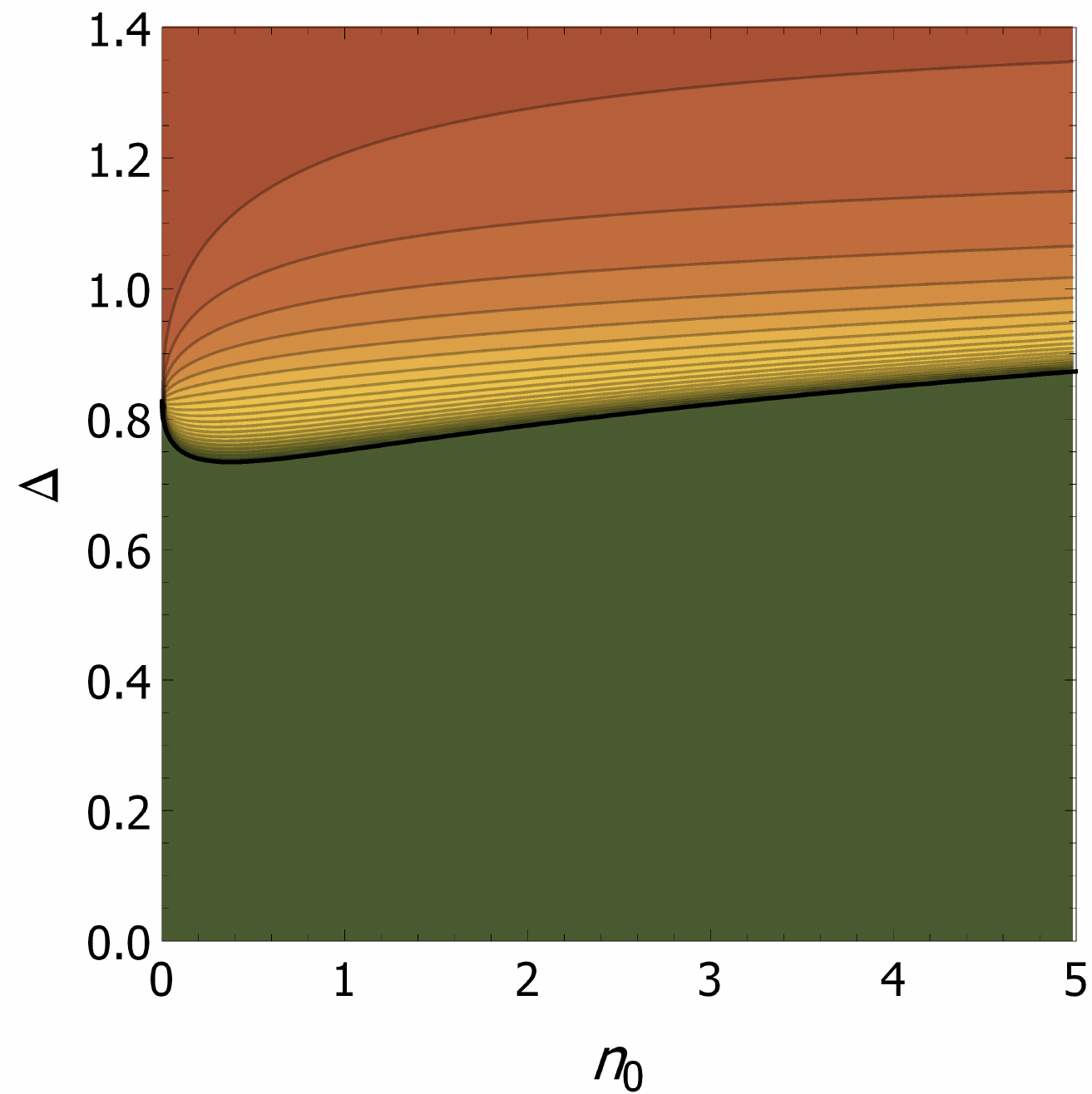} & \includegraphics[height=6.5cm]{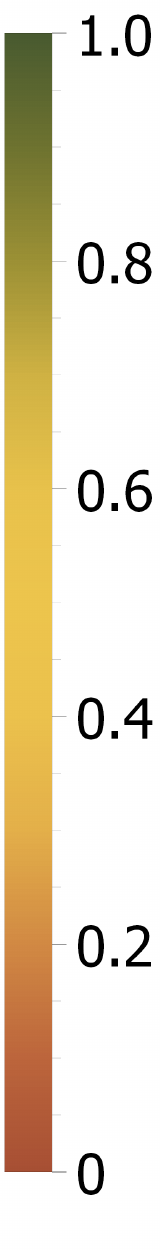}\tabularnewline
\end{tabular}
\par\end{centering}
\caption{\label{f:betaOpt_RLD}Optimal squeezing fraction
$\beta_{opt}$ as function of the noise parameter $\Delta$ and number
of photons $n_0$ in the probe. The threshold $\Delta_{c}\left(n_0\right)$
is represented with a black curve.}
\end{figure}

\begin{figure}[H]
\begin{centering}
{\bf (a)}\hskip4cm{\bf (b)}\\
\begin{tabular}{cc}
\includegraphics[width=0.50\columnwidth]{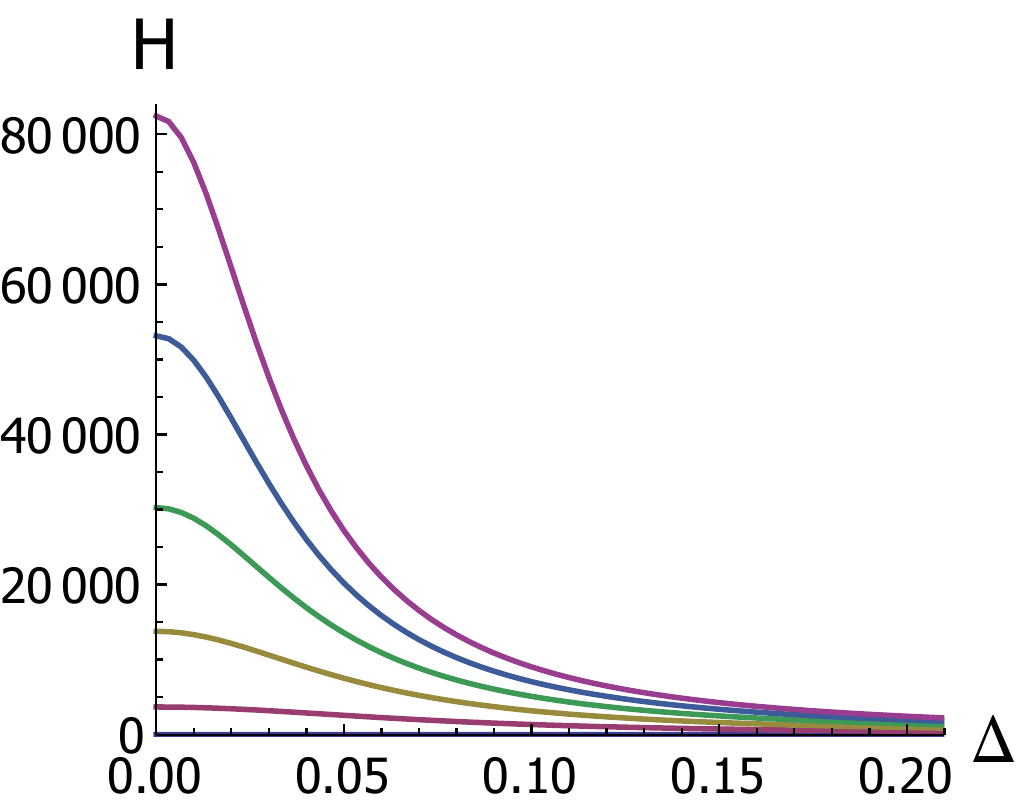} &
\includegraphics[width=0.48\columnwidth]{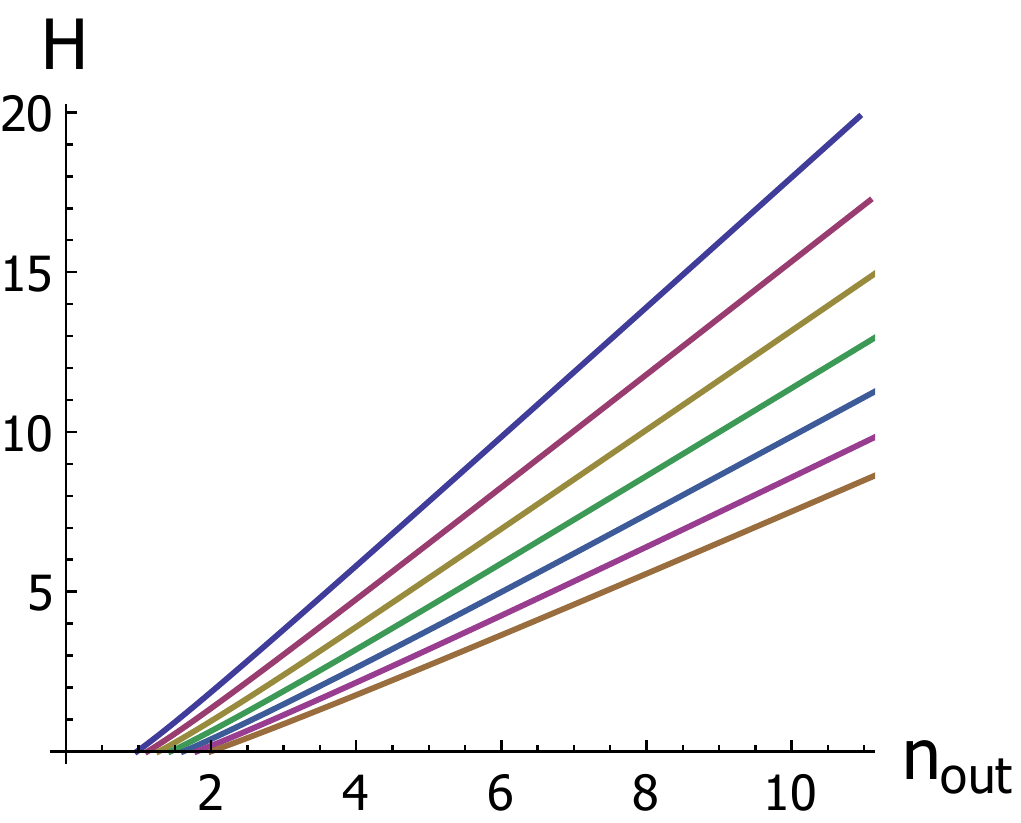}
\tabularnewline
\end{tabular}
\par\end{centering}
\caption{\label{f:QFI_RLD}(a) Optimized QFI in the case of 
random linear disturbance, as function of the noise
parameter $\Delta$ for fixed values of the average photon number; from bottom-left to top-right
$n_0=\{1,\,21,\,41,\,61,\,81,\,101\}$. \\
(b) Optimized QFI as a function of the output photon number
$n_{\sf out}$ for fixed values of the noise parameter $\Delta$; 
from top to bottom $\Delta=\{0.7,\,0.75,\,0.8,\,0.85,\,0.9,\,0.95,\,1.0\}$.}
\end{figure}

We now turn our attention to a situation
where the disturbance parameter is not fixed, 
being a random variable satisfying a (known) probability 
distribution. To keep the situation as  general
and symmetric as possible, we consider two 
disturbance operators in the Hamiltonian, leading to a
displacement in phase-space along orthogonal directions.
In formula, we consider the unitary evolution
\begin{align}
U_{\phi,\boldsymbol{\eta}} = \exp \{-i(\phi\: a^\dag a + \eta_1 Q+ \eta_2 P )\} \:,
\end{align}
where $\boldsymbol{\eta}=(\eta_1,\eta_2)^{\sf T}$.
If we consider the two disturbance parameters both distributed according to a Gaussian 
probability distribution centered at zero with the same variance $\Delta$, the
average output (mixed) state reads,
\begin{equation}
\mathcal{G}_{\phi,\Delta} (|\psi_0\rangle\langle\psi_0|)=\int_{\mathbb{R}^{2}}\mbox{d}\eta_{1} \mbox{d}\eta_{2}\:\frac{e^{-\frac{\eta_{1}^{2}+\eta_{2}^{2}}{2\Delta^{2}}}}{2\pi\Delta^{2}}U_{\phi,\boldsymbol{\eta}}\: |\psi_0\rangle\langle\psi_0|\: U_{\phi,\boldsymbol{\eta}}^{\dagger}\label{eq:Random-displacement} \:,
\end{equation}
We still consider as an input a generic pure Gaussian state, $|\psi_0\rangle = D(\alpha) S(\xi) |0\rangle$.
In absence of the additional phase-rotation, this channel is usually referred to as 
{\em Gaussian noise} \cite{GaussAOP}. As its effect is to displace the state incoherently in different
directions of phase space with random amplitude, the output state will become 
mixed similar to a state in a phase-diffusion channel. However we can
observe at least two main differences between this noisy channel and phase-diffusion: the latter 
corresponds to a random phase-rotation, {\em i.e.}  one has a single disturbance operator 
$A=a^\dag a$; as a consequence the action of the channel commutes with the phase-rotation
itself, and one can consider the two evolutions separately. Moreover, the state after
a phase-diffusion channel will have the same amount of mean photon number
of the input state, while after the channel described in Eq. (\ref{eq:Random-displacement}),
the output state will have an output average photon number $n_{\sf out}=n_0 + 2\Delta^2$.
It is straightforward to see that
the state remains Gaussian and, consequently, one can evaluate the QFI by using
the formulas presented in Sec. \ref{s:estiGauss}. As in the {\em unitary disturbance}
case, we focus on small fluctuations around the value $\phi=0$, and we
optimize the input Gaussian state for a fixed number of photons
$n_0$, over the squeezing angle $\theta$ and the squeezing fraction $\beta$.
If the optimal probe has a nonzero displacement
parameter $\alpha$, the optimal squeezing angle is $\theta_{opt}=\pi$
(phase squeezing), while, due to the symmetry of the disturbance introduced,
the QFI does not depend on $\theta$ when the state is prepared in a 
squeezed vacuum state. We now discuss the behavior of the optimal squeezing 
fraction $\beta_{\sf opt}$ for a fixed average input photon
number $n_0$, and varying the disturbance parameter $\Delta$, 
by observing the contour plot of Fig. \ref{f:betaOpt_RLD}. For 
values of $\Delta$ smaller than a threshold $\Delta_{t}\left(n_0\right)$
(the black line in Fig. \ref{f:betaOpt_RLD}), the optimal state is 
always a squeezed vacuum state, and we have $\beta_{\sf opt}=1$ as 
in the noiseless case. By increasing $\Delta$,  
the optimal squeezing fraction drops to $\beta_{\sf opt}=0$: 
in more detail, for very small values of $n_0$, when
$\Delta$ crosses the limiting value $\Delta_t(0)=\frac{\sqrt{1+\sqrt{3}}}{2}$ 
the optimal state changes
very abruptly from a squeezed vacuum to a coherent state.
For larger values of $n_0$, there is a region in the parameter space $(n_0,\Delta)$
where the optimal state is a displaced squeezed vacuum state.
By studying more carefully the noise threshold, we observe that
it has a minimum $\Delta_{t,min}\approx0.734$
at $n_0\approx0.375$, below which the optimal probe is the squeezed
vacuum state irrespective of the number of photons in the probe. 
Moreover the corresponding function can be well aproximated for $n_0\gtrsim 20$, as 
$\Delta_{t}\left(n_0\right)\thicksim\sqrt[6]{n_0/16}$, showing how the
threshold increases very slowly with the mean photon number.
Remarkably, one also observes that for a fixed $\Delta$, for large values of the
input energy $n_0$ the optimal state is still a squeezed vacuum state, suggesting
that squeezing is still a resource when the noise parameter is 
small compared to the mean photon number. 
The behavior of the optmized QFI as a function of the noise parameter $\Delta$ 
is displayed in Fig. \ref{f:QFI_RLD} (a). Despite the fact that
the energy of the output state increases for nonzero noise
parameter, the ultimate estimation performances are degraded by the {\em random linear
disturbance}. As in the previous case, we also plot the behavior of the
QFI as a function of the output photon number $n_{\sf out}$ in Fig. \ref{f:QFI_RLD} (b).
We clearly observe that in this case the quadratic behavior is lost, and that the QFI
scales linearly with $n_{\sf out}$. Remarkably we can also compute
an approximate value for the linear coefficient, obtaining 
\begin{align}
H \approx a \: n_{\sf out} \:, \:\:\:\textrm{with}\:\: a \approx \frac{1}{\Delta^2} \:.
\end{align}
The approximation is more accurate for large values of the 
photon number, where the optimal probe state is more likely to be
in a squeezed vacuum state. 
As expected, smaller values of $\Delta$ correspond to larger values of the coefficient
and thus to larger values of the QFI; However we observe that any
non-zero values of the noise parameter are enough to lose the non-classical scaling. This 
discontinuity in the scaling of the QFI is quite typical in noisy quantum metrology as
it has been indeed widely observed in many different metrological problems \cite{Rafal09,Rafal12}.
\subsection*{Performance of homodyne detection}
\begin{figure}
\begin{centering}
\includegraphics[width=0.8\columnwidth]{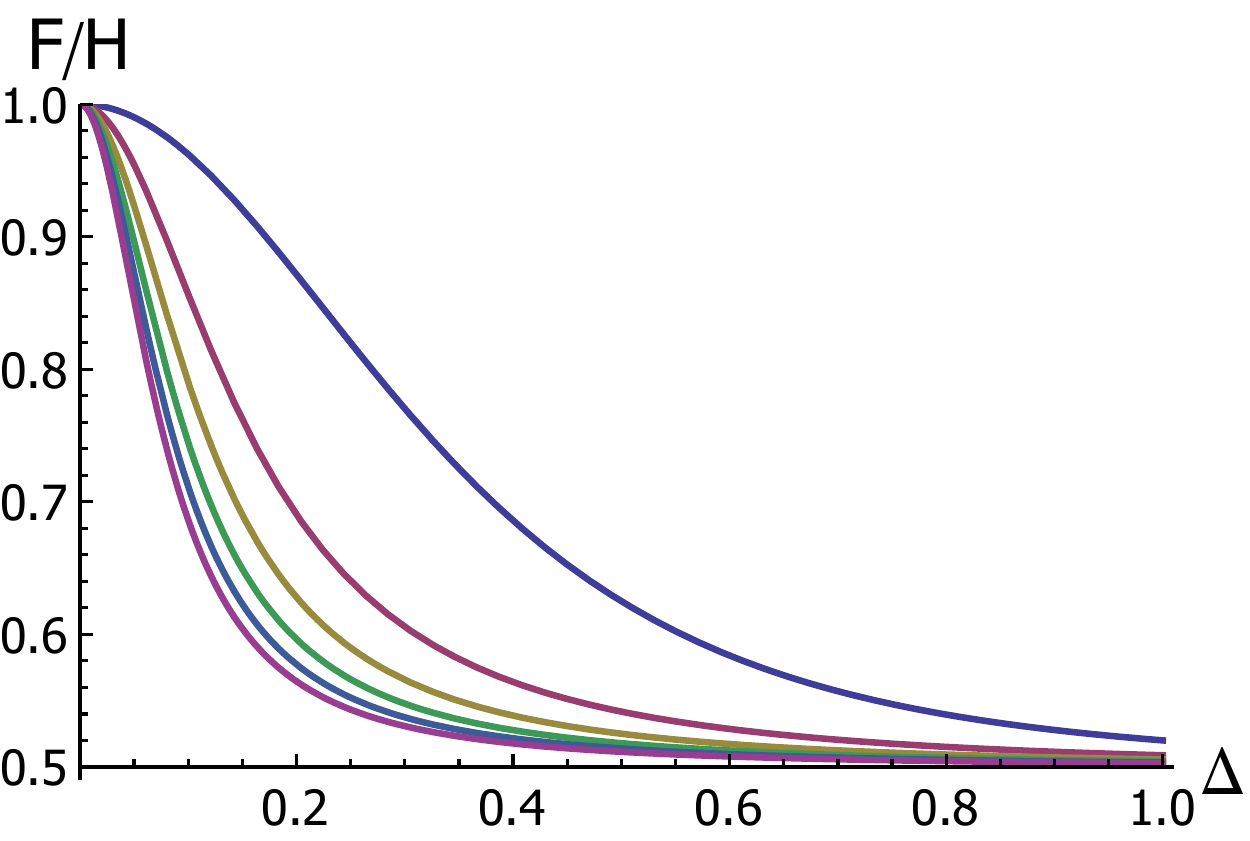}
\end{centering}
\caption{\label{f:RLD_Homo_Sq}
(Color online) Ratio between
the Fisher Information of the homodyne detection and the corresponding
QFI for squeezed vacuum input states undergone random linear disturbance. The ratio is plotted
as a function of the noise parameter $\Delta$ and for different
values of the average number of photons of the probe.
From top to bottom: $n_0=\{10^{-9},2,4,6,8,10\}$.}
\end{figure}

\begin{figure}
\begin{centering}
\begin{tabular}{cc}
\includegraphics[width=0.75\columnwidth]{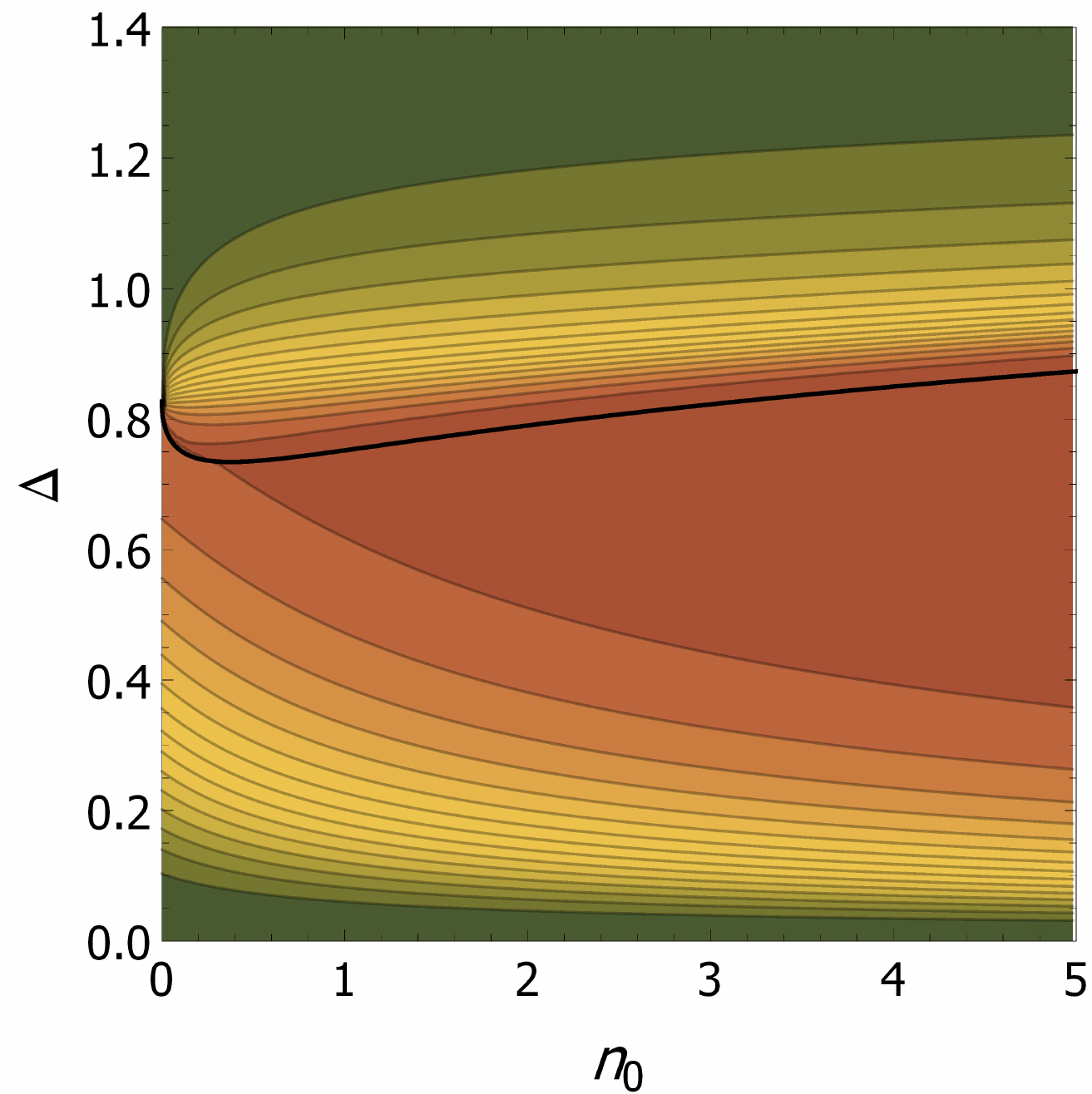} & 
\includegraphics[width=0.1\columnwidth]{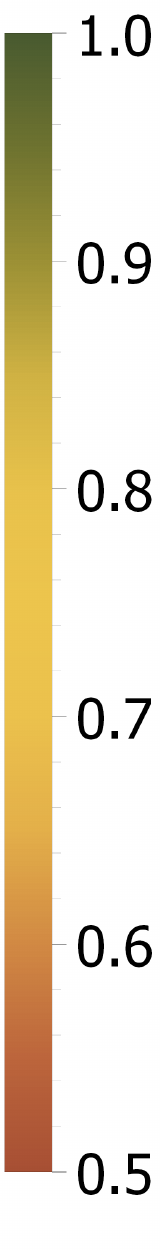}
\tabularnewline
\end{tabular}
\par\end{centering}
\caption{\label{f:RLD_Homo_Optimal} (Color online) Ratio $F/H$ between the homodyne's
FI over the QFI for optimized input state as a function of the input photon number $n_0$ and
of the noise parameter $\Delta$. The superimposed black line corresponds to the input state threshold 
$\Delta_t(n_0)$ which divides regions with squeezed vacuum probes ($\Delta<\Delta_t(n_0)$) 
and displaced squeezed probes ($\Delta> \Delta_t(n_0)$).}
\end{figure}

As in the previous section we discuss the efficiency of homodyne
detection for a 
random linear disturbance channel. The corresponding FI can be 
calculated by starting from the conditional probability 
\begin{align}
p(x|\phi) = \langle x_\omega | \: \mathcal{G}_{\phi,\Delta}(|\psi_0\rangle\langle\psi_0|)\: | x_\omega\rangle \:.
\end{align} 
As before, we optimize over the homodyne angle $\omega$ and
evaluate the ratio $F/H$ between the optimized FI and the QFI. \\
Before focusing on the optimal input states identified in the previous section, 
we study the optimality of homodyne detection for the two extreme cases,
that is for input squeezed vacuum states ($\beta=1$) and coherent states
($\beta=0$). Regarding the first case, {\em i.e.} $|\psi_0\rangle = S(r)|0\rangle$, 
the ratio $F/H$ is plotted in Fig. \ref{f:RLD_Homo_Sq}. If we fix the 
value of the input photon number $n_0=\sinh^2 r$, homodyne detection
results to be (nearly) optimal only for small values of the noise parameter
$\Delta$, and the ratio $F/H$ decreases monotonically to the asymptotic
value $F/H \approx 0.5$ by increasing $\Delta$.
If we rather consider coherent states as the input, one can easily obtain
that homodyne detection is always optimal, for every value of the 
input energy and the noise parameter $\Delta$.\\
Having in mind these results for the extreme cases, we can now discuss
the optimality of homodyne detection where we consider the optimal
input state for each value of the noise parameter $\Delta$ and of the 
input average photon number $n_0$.
It is useful to compare the corresponding 
ratio $F/H$, plotted in Fig. \ref{f:RLD_Homo_Optimal},
with the plot of the optimal squeezing fraction
in Fig. \ref{f:betaOpt_RLD}, that we have discussed previously in this section.
If we fix the value of the input photon number $n_0$, and vary
the noise parameter $\Delta$, we observe the following behavior:
the homodyne measurement is optimal for $\Delta \approx 0$, and its efficiency decreases reaching 
a minimum for a certain value of $\Delta$. The ratio $F/H$ starts to increase
with $\Delta$, reaching values near to unity for large values of $\Delta$.
In particular we observe that the minimum of the ratio $F/H$ occurs in proximity
of the threshold $\Delta_t(n_0)$ described in the previous section, which divides
the different regions of parameters for the optimal squeezing fraction. In Fig. 
\ref{f:RLD_Homo_Optimal}, $\Delta_t(n_0)$ is depicted as a superimposed black
line. In the region below the black line ($\Delta < \Delta_t(n_0)$), 
the optimal input state is a squeezed
vacuum, and the behavior of the ratio $F/H$ does indeed correspond
to the plots in Fig. \ref{f:RLD_Homo_Sq}. In the region above the black
line ($\Delta>\Delta_t(n_0)$), 
the optimal state is a displaced squeezed state, which tends to a 
coherent state for larger noise. As a consequence 
the ratio $F/H$ starts to increase by increasing $\Delta$, reaching 
eventually the {\em optimality}, which, as discussed above, is always
obtained for input coherent states. 
\section{Conclusions} \label{s:concl}
The effect of noise and imperfection on the performances of quantum metrology protocols
has received a lot of attention in the recent years. In this paper, we have discussed the case
where an unwanted term is present in the Hamiltonian generating the phase-shift that one 
wants to estimate. In particular we have considered both unitary and random linear disturbance, 
with input Gaussian state, optimizing over the squeezing fraction and the squeezing angle. 
While in the case of {\em unitary disturbance} squeezed vacuum is shown to be the optimal probe state and the non-classical quadratic scaling of the QFI is still observed, in the presence of {\em random disturbance} the optimal squeezing fraction crucially depends on the input energy and on the noise parameter values, and, more importantly, any non-zero value of the noise is enough to cause a linear scaling between the QFI and the number of photons of the quantum state. \\
We have also discussed the performance of homodyne detection, which is shown to be in general an efficient measurement in both cases, despite the fact that the optimality is observed only in some regions
of the parameter space characterizing the input state and the noisy channel. 
\section{Acknowledgments} 
DDS acknowledges support from the Brazilian funding agencies CNPq, through the grant 237095/2012-2, and FAPESP, through the grant 2011/00220-5. DSS is also thankful to the CQD theory group at Imperial College London for the warm hospitality. 
MGG acknowledges support from EPSRC through grant EP/K026267/1.
MSK thanks the UK EPSRC for financial support.
\bibliography{MyLibrary}

\end{document}